\documentclass[aps,pra,twocolumn,groupedaddress,nofootinbib]{revtex4-1}
\usepackage{amsmath, amsfonts, amssymb,amsthm}
\usepackage{graphicx,subfigure}
\usepackage{mathrsfs}
\usepackage[mathscr]{eucal}
\usepackage{times,txfonts}
\usepackage{physics}
\usepackage{bm}
\usepackage{subfigure}
\usepackage[bookmarks=false]{hyperref}
\usepackage{xcolor}
\usepackage{makecell}
\hypersetup{colorlinks=true,citecolor=blue,
linkcolor=blue,urlcolor=blue,pdfstartview=FitH,
bookmarksopen=true}

\usepackage{color,soul}

\begin{document}
\title{Maximal steered coherence in the background of Schwarzschild space-time}
\author{Ming-Ming Du$^1$}
\email{mingmingdu@njupt.edu.cn}
\author{Hong-Wei Li$^1$, Shu-Ting Shen$^1$, Xiao-Jing Yan$^2$, Xi-Yun Li$^2$, Lan Zhou$^2$,Wei Zhong$^3$,}
\author{Yu-Bo Sheng$^{1,3}$}
\email{shengyb@njupt.edu.cn}

\affiliation{$1.$ College of Electronic and Optical Engineering and College of Flexible Electronics (Future Technology), Nanjing
University of Posts and Telecommunications, Nanjing, 210023, China\\
$2.$ School of Science, Nanjing University of Posts and Telecommunications, Nanjing,
210023, China\\
$3.$ Institute of Quantum Information and Technology, Nanjing University of Posts and Telecommunications, Nanjing, 210003, China}
\date{\today}
\begin{abstract}
In the past two decades, the exploration of quantumness within Schwarzschild spacetime has garnered significant interest, particularly regarding the Hawking radiation's impact on quantum correlations and quantum coherence. Building on this foundation, we investigate how Hawking radiation influences maximal steered coherence (MSC)—a crucial measure for gauging the ability to generate coherence through steering. We find that as the Hawking temperature increases, the physically accessible MSC degrade while the unaccessible MSC increase. This observation is attributed to a redistribution of the initial quantum correlations, previously acknowledged by inertial observers, across all bipartite modes. In particular, we find that in limit case that the Hawking temperature tends to infinity, the accessible MSC equals to $1/\sqrt{2}$ of its initial value, and the unaccessible MSC between mode $A$ and $\overline{B}$ also equals to the same value. Our findings illuminate the intricate dynamics of quantum information in the vicinity of black holes, suggesting that Hawking radiation plays a pivotal role in reshaping the landscape of quantum coherence and entanglement in curved spacetime. This study not only advances our theoretical understanding of black hole thermodynamics but also opens new avenues for investigating the interface between quantum mechanics and general relativity.
\end{abstract}

\maketitle

\section{Introduction}
In the past two decades, a burgeoning body of literature has emerged at the intersection of general relativity and quantum information science, focusing on the exploration of quantum entanglement\cite{MartinMartinez2010,Wang2010a,FuentesSchuller2005,Alsing2006,Adesso2007,Xu2014,Dai2016,Hwang2011,TorresArenas2019,MartinMartinez2011,Bruschi2012,Dong2018,Dong2020,Qiang2018,Wu2022,Wu2022a,Wu2024,Wu2023a,Jing2023,He2018a,Qiang2015,Deng2011}, quantum discord\cite{Wang2010b,Wu2023a,Qiang2015}, quantum coherence\cite{Wu2021,Tian2012,He2024,He2018}, and quantum steering\cite{Wu2023a,Wang2016,Wu2023a} within the framework of general relativity. This interdisciplinary research endeavor seeks to unravel the intricate behavior of quantum phenomena in the presence of gravitational fields, particularly around phenomena such as black holes, where the interplay between quantum mechanics and general relativity becomes most pronounced.

The study of quantum entanglement\cite{MartinMartinez2010,Wang2010a,FuentesSchuller2005,Alsing2006,Adesso2007,Xu2014,Dai2016,Hwang2011,TorresArenas2019,MartinMartinez2011,Bruschi2012,Dong2018,Dong2020,Qiang2018,Wu2022,Wu2022a,Wu2024,Wu2023a,Jing2023,He2018a,Qiang2015,Deng2011} in relativistic settings laid the foundational stone, illuminating how entangled particles behave when subjected to the curved space-time around massive objects. Following this, research expanded into the quantum discord\cite{Wang2010b,Wu2023a,Qiang2015}, which offering insights into the quantum correlations present in mixed states that go beyond entanglement\cite{Ollivier2001}. Quantum coherence\cite{Baumgratz2014}, a measure of the superposition of quantum states, and quantum steering\cite{Wiseman2007}, a phenomenon whereby one system can nonlocally influence another's state through quantum nolocality, have also been investigated under these extreme conditions\cite{Wu2021,Tian2012,He2024,He2018,Wu2023a,Wang2016,Wu2023a}.

These studies are crucial steps towards understanding the fundamental nature of the universe. They bridge the gap between the macroscopic predictions of general relativity and the microscopic descriptions of quantum mechanics, with black holes serving as a natural laboratory for such investigations. Specifically, the phenomenon of Hawking radiation from black holes provides a unique setting to examine these quantum effects under the influence of strong gravitational fields.

Our work builds upon this rich background, focusing on how Hawking radiation affects maximal steered coherence (MSC), a measure that quantifies the ability to generate quantum coherence remotely through steering. We find that as the Hawking temperature increases, there is a degradation in the physically accessible MSC, while the inaccessible MSC increases. This suggests that the initial quantum correlations, as described by inertial observers, undergo redistribution among all bipartite modes due to the influence of Hawking radiation. Through this analysis, we contribute to the ongoing dialogue on the quantum characteristics of the universe under the influence of gravity, aiming to further the understanding of how quantum information processes are affected by and can probe the structure of spacetime.

To be self-contained, we organize the rest of this paper as follows. In Sec. \ref{sec1}, we give a brief introduction of MSC. In Sec. \ref{sec2}, we briefly introduce the quantization of
Dirac fields in a Schwarzschild black hole. In Sec. \ref{sec3}, we investigate the effect of Hawking radiation for the maximal steered coherence.  Finally, we conclude this work in Sec. \ref{sec5}.

\section{The maximal steered coherence}\label{sec1}
In this section, we briefly introduce the concept of maximal steered coherence (MSC).
In our exploration of maximal steered coherence (MSC), we consider an intriguing scenario involving two participants: Alice ($A$) and Bob ($B$), who shared a two-qubit state, denoted as $\rho_{AB}$. Initially, Alice performs the Positive Operator-Valued Measure (POVM) measurements, $M$, on her qubit ($A$), leading to the consequential state collapse of Bob's qubit ($B$) to:
\begin{align}\label{du1}
\rho_{B|M} = \frac{\mathrm{tr}_A(M \otimes \mathbb{I})\rho_{AB}}{p_M},
\end{align}
where $p_M = \mathrm{tr}(M \otimes \mathbb{I} \rho_{AB})$ represents the probability of achieving the state $\rho_{B|M}$. To effectively quantify the resultant coherence on Bob's qubit, a reference basis $\{\lvert\xi_i\rangle\}$, composed of the eigenbasis of $\rho_B = \mathrm{tr}_A \rho_{AB}$, is selected.
The maximal steered coherence (MSC) accessible by Bob can be expressed as\cite{Hu2016}:
\begin{align}\label{du2}
C(\rho_{AB}) = \inf_{\{\lvert\xi_i\rangle\}} \left\{ \max_{M \in \mathrm{POVM}} C_{l_1}^{\{\lvert\xi_i\rangle\}}(\rho_{B|M}) \right\},
\end{align}
integrating an infimum over $\{\lvert\xi_i\rangle\}$ to accommodate for the potential degeneracy of $\rho_{B}$, and $C_{l_1}(\rho_{B|M})$ measuring the coherence present within $\rho_{B|M}$.

\section{Vacuum structure and Hawking radiation for Dirac fields}\label{sec2}
In this section, we briefly introduce the quantization of Dirac fields in a Schwarzschild black
hole. We consider a Schwarzschild black hole, characterized by the metric\cite{MartinMartinez2010}:
\begin{equation}\label{du3}
ds^2 = -\left(1-\frac{2M}{r}\right)dt^2 + \left(1-\frac{2M}{r}\right)^{-1}dr^2 + r^2(d\theta^2 + \sin^2\theta d\varphi^2),
\end{equation}
where $r$ and $M$ denote the radial coordinate and black hole mass, respectively.  For simplicity, we set Planck's constant ($\hbar$), gravitational constant ($G$), speed of light ($c$), and Boltzmann's constant ($k$) to unity. The Dirac equation in this spacetime can be expressed as
\begin{align}
&-\frac{\gamma_0}{\sqrt{1-\frac{2M}{r}}}\frac{\partial\Phi}{\partial t} + \gamma_1\sqrt{1-\frac{2M}{r}}\left[\frac{\partial}{\partial r} + \frac{1}{r} + \frac{M}{2r(r-2M)}\right]\Phi \\\notag
 &+\frac{\gamma_2}{r}\left(\frac{\partial}{\partial\theta} + \frac{\cot\theta}{2}\right)\Phi + \frac{\gamma_3}{r\sin\theta}\frac{\partial\Phi}{\partial\varphi} = 0,
\end{align}
with $\gamma_i (i = 0, 1, 2, 3)$ as the Dirac matrices\cite{Jing2004,Wang2010}.
Solving the Dirac equation near the event horizon of black hole, a set of positive (fermions) frequency outgoing solutions inside and outside regions of the event horizon can be obtained
\begin{equation}\label{du6}
\Phi_{\mathbf{k},\mathrm{in}}^{+} \sim \phi(r)e^{i\omega u},
\end{equation}
and
\begin{equation}\label{du7}
\Phi_{\mathbf{k},\mathrm{out}}^{+} \sim \phi(r)e^{-i\omega u},
\end{equation}
where $\phi(r)$ denotes a four-component Dirac spinor. The variable $u = t - r_*$, with $r_* = r + 2M\ln\left(\frac{r - 2M}{2M}\right)$ being the tortoise coordinate. The wave vector $\mathbf{k}$ and the frequency $\omega$ are related by the dispersion relation $|\mathbf{k}| = \omega$ for the massless Dirac field. The Dirac field $\Phi$ in the vicinity of a Schwarzschild black hole can be expanded in terms of the quantum field operators as follows:
\begin{equation}
\begin{aligned}
\Phi=\int d\mathbf{k} \bigg[ &\hat{a}_{\mathbf{k}}^{\text{in}} \Phi_{\mathbf{k},\text{in}}^{+} + \hat{b}_{\mathbf{k}}^{\text{in}\dagger} \Phi_{\mathbf{k},\text{in}}^{-}+\hat{a}_{\mathbf{k}}^{\text{out}} \Phi_{\mathbf{k},\text{out}}^{+} + \hat{b}_{\mathbf{k}}^{\text{out}\dagger} \Phi_{\mathbf{k},\text{out}}^{-} \bigg],
\end{aligned}
\end{equation}
where $\hat{a}_\mathbf{k}^{\text{in}}$ and $\hat{b}_\mathbf{k}^{\text{in}\dagger}$ represent the annihilation and creation operators for fermions and antifermions within the event horizon, respectively. Similarly, $\hat{a}_\mathbf{k}^{\text{out}}$ and $\hat{b}_\mathbf{k}^{\text{out}\dagger}$ denote the annihilation and creation operators for the exterior region. These operators satisfy the canonical anticommutation relations:
\begin{equation}
\{ \hat{a}_{\mathbf{k}}^{\text{out}}, \hat{a}_{\mathbf{k}'}^{\text{out}\dagger} \} = \{ \hat{b}_{\mathbf{k}}^{\text{in}}, \hat{b}_{\mathbf{k}'}^{\text{in}\dagger} \} = \delta(\mathbf{k} - \mathbf{k}').
\end{equation}
The Schwarzschild vacuum state is thus defined by $\hat{a}_{\mathbf{k}}^{\text{in}} |0\rangle_S = \hat{a}_{\mathbf{k}}^{\text{out}} |0\rangle_S = 0$. Consequently, the solutions $\Phi_{\mathbf{k},\text{in}}^{\pm}$ and $\Phi_{\mathbf{k},\text{out}}^{\pm}$ are commonly referred to as Schwarzschild modes.

Accordingly, the Dirac fields in Kruskal spacetime can be expanded utilizing these Kruskal modes:

\begin{align}
\Phi &= \int d\mathbf{k} [2\cosh(4\pi M \omega)]^{-\frac{1}{2}} \left[\hat{c}_{\mathbf{k}}^{\mathrm{in}} \Psi_{\mathbf{k},\mathrm{in}}^{+} + \hat{d}_{\mathbf{k}}^{\mathrm{in}} \Psi_{\mathbf{k},\mathrm{in}}^{-} \right. \\
&\quad \left. + \hat{c}_{\mathbf{k}}^{\mathrm{out}} \Psi_{\mathbf{k},\mathrm{out}}^{+} + \hat{d}_{\mathbf{k}}^{\mathrm{out}\dagger} \Psi_{\mathbf{k},\mathrm{out}}^{-} \right],
\end{align}
where $\hat{c}_\mathbf{k}^{\sigma}$ and $\hat{d}_\mathbf{k}^{\sigma\dagger}$, with $\sigma = (\text{in}, \text{out})$, denote the fermion annihilation and antifermion creation operators acting on the Kruskal vacuum.

Adopting the methodology proposed by Domour and Ruffini facilitates\cite{Damour1976} the establishment of a complete set of positive energy modes, namely, the Kruskal modes. This formulation enables a seamless connection between Eq. (\ref{du6}) and (\ref{du7}), enhancing our understanding of quantum field behavior in curved spacetime. The Kruskal modes are expressed as:
\begin{gather*}
\Psi_{\mathbf{k},\mathrm{out}}^{+} = e^{-2\pi M \omega} \Phi_{-\mathbf{k},\mathrm{in}}^{-} + e^{2\pi M \omega} \Phi_{\mathbf{k},\mathrm{out}}^{+}, \\
\Psi_{\mathbf{k},\mathrm{in}}^{+} = e^{-2\pi M \omega} \Phi_{-\mathbf{k},\mathrm{out}}^{-} + e^{2\pi M \omega} \Phi_{\mathbf{k},\mathrm{in}}^{+}.
\end{gather*}

Utilizing Kruskal modes, the Dirac fields in Kruskal spacetime can be expanded as follows:
\begin{align}
\Phi &= \int d\mathbf{k} \left[2\cosh(4\pi M\omega)\right]^{-\frac{1}{2}} [\hat{c}_{\mathbf{k}}^{\text{in}} \Psi_{\mathbf{k},\text{in}}^{+} + \hat{d}_{\mathbf{k}}^{\text{in}} \Psi_{\mathbf{k},\text{in}}^{-} \\\notag
&+ \hat{c}_{\mathbf{k}}^{\text{out}} \Psi_{\mathbf{k},\text{out}}^{+}
+ \hat{d}_{\mathbf{k}}^{\text{out}\dagger} \Psi_{\mathbf{k},\text{out}}^{-}],
\end{align}
where $\hat{c}_\mathbf{k}^{\sigma}$ and $\hat{d}_\mathbf{k}^{\sigma\dagger}$, with $\sigma = (\text{in}, \text{out})$, are the fermion annihilation and antifermion creation operators acting on the Kruskal vacuum.
Equations (25) and (28) delineate disparate decompositions of the identical Dirac field within Schwarzschild and Kruskal frameworks, respectively. These decompositions lead to the Bogoliubov transformation connecting Kruskal and Schwarzschild operators, as demonstrated below:

\begin{equation}
\begin{aligned}
\hat{c}_\mathbf{k}^{\text{out}} &= \frac{1}{\sqrt{e^{-8\pi M \omega} + 1}} \hat{a}_\mathbf{k}^{\text{out}} - \frac{1}{\sqrt{e^{8\pi M \omega} + 1}} \hat{b}_\mathbf{k}^{\text{out}\dagger}, \\
\hat{c}_\mathbf{k}^{\text{out}\dagger} &= \frac{1}{\sqrt{e^{-8\pi M \omega} + 1}} \hat{a}_\mathbf{k}^{\text{out}\dagger} - \frac{1}{\sqrt{e^{8\pi M \omega} + 1}} \hat{b}_\mathbf{k}^{\text{out}}.
\end{aligned}
\end{equation}

Thus, the Kruskal vacuum and excited states can be succinctly represented within the Schwarzschild Fock space framework as follows:
\begin{equation}\label{du4}
\begin{aligned}
|0\rangle_K &= C |0\rangle_{\text{out}} |0\rangle_{\text{in}} +S|1\rangle_{\text{out}} |1\rangle_{\text{in}}, \\
|1\rangle_K &= |1\rangle_{\text{out}} |0\rangle_{\text{in}},
\end{aligned}
\end{equation}
where $T =1/(8\pi M)$ denotes the Hawking temperature, $C=1/\sqrt{e^{-\frac{\omega}{T}}+1}$ and $S=1/\sqrt{e^{\frac{\omega}{T}}+1}$. Here, $\{|n\rangle_{\text{out}}\}$ and $\{|n\rangle_{\text{in}}\}$ respectively represent the Schwarzschild number states for fermions outside the event horizon and antifermions within.

\section{The maximal steered coherence in Schwarzschild spacetime}\label{sec3}
Consider two maximally entangled fermionic modes in the asymptotically flat region of the Schwarzschild black hole:
\begin{equation}\label{du5}
|\phi_{AB}\rangle = \frac{1}{\sqrt{2}}\left(|0\rangle_A |1\rangle_B - |1\rangle_A |0\rangle_B\right),
\end{equation}
where the subscripts $A$ and $B$ denote the modes associated with observers Alice and Bob, respectively. After their coincidence, Alice remains stationary in the asymptotically flat region, while Bob hovers outside the event horizon. Bob's detection reveals a thermal Fermi-Dirac distribution of fermions. Utilizing Eq. (\ref{du4}), we can rewritten Eq. (\ref{du5}) in terms of Kruskal modes for Alice and Schwarzschild modes for Bob as:
\begin{equation}
\begin{aligned}
|\phi_{AB\bar{B}}\rangle &= \frac{1}{\sqrt{2}}\left(|0\rangle_A|1\rangle_B|0\rangle_{\bar{B}}-C|1\rangle_A|0\rangle_B|0\rangle_{\bar{B}} \right. \\
&\quad - \left. S|1\rangle_A|1\rangle_B|1\rangle_{\bar{B}} \right),
\end{aligned}
\end{equation}
where $\bar{B}$ represents the mode observed by a hypothetical observer, Anti-Bob, within the event horizon.

Since the exterior region is causally disconnected from the interior region of the black hole, the only information which is physically accessible to the observers is encoded in the mode $A$ described by Alice and the mode $B$ outside the event horizon described by Bob. Taking the trace over the $B$ mode inside the event horizon, we can obtain
\begin{equation}\label{rab}
\rho_{AB}=\left(
  \begin{array}{cccc}
    0 & 0 & 0 & 0 \\
    0 & \frac{1}{2} & -\frac{1}{2}C & 0 \\
    0 & -\frac{1}{2}C & \frac{1}{2}C^2 & 0 \\
    0 & 0 & 0 & -\frac{1}{2}S^2 \\
  \end{array}
\right)
\end{equation}

To calculate MSC of $\rho_{AB}$, we needs to take the maximization over the set of projective measurements $M= (1+ \vec{m} \cdot \sigma)/2$, where $\sigma= (\sigma_{x}, \sigma_{y}, \sigma_{z})$ and $\vec{m} = (\sin \theta\cos \phi, \sin \theta\sin \phi, \cos \theta)$, with $\theta$ and $\phi$ being the polar and azimuth angles, respectively. Then, the post-measurement state of qubit $B$ can be obtained as
\begin{align}
\rho_{B|M}=\left(
\begin{array}{cc}
 -\frac{C^2 (\cos (\theta)-1)}{C^2+S^2-\left(C^2+S^2-1\right) \cos (\theta)+1} & -\frac{e^{-i \phi} C \sin (\theta)}{C^2+S^2-\left(C^2+S^2-1\right) \cos (\theta)+1} \\
 -\frac{e^{i \phi} C \sin (\theta)}{C^2+S^2-\left(C^2+S^2-1\right) \cos (\theta)+1} & -\frac{-S^2+\left(S^2-1\right) \cos (\theta)-1}{C^2+S^2-\left(C^2+S^2-1\right) \cos (\theta)+1} \\
\end{array}
\right)
\end{align}
with $p_M=[1+C^2+S^2-(C^2+S^2-1)\cos(\theta)]/2$.
So the MSC can be obtained as
\begin{align}
C(\rho_{AB})=\frac{1}{\sqrt{1+e^{-\frac{1}{T}}}},
\end{align}
when $\theta=\arccos[(C^2+S^2-1)/(1+C^2+S^2)]$ and $\phi$ can take any value.
We can see that the MSC depends on the Hawking temperature $T$, i.e., Hawking radiation of the black hole influences the MSC.

\begin{figure}
\includegraphics[width=0.44\textwidth]{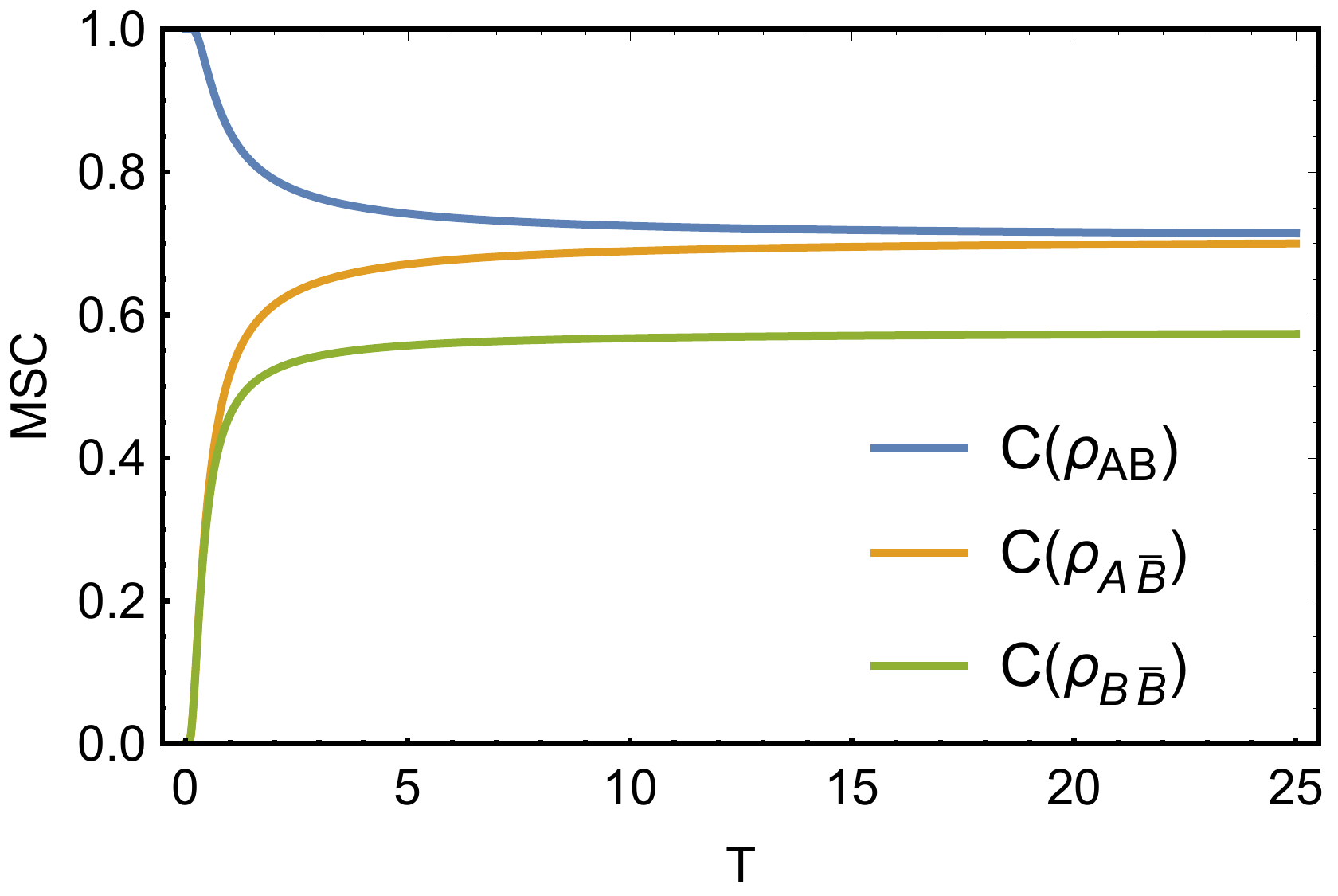}
\caption{The MSC as functions of the Hawking temperature $T$ for $\omega=1$. Here, the unit of temperature is presented by the natural unit system, wherein the Planck constant \(\hbar\), gravitational constant \(G\), speed of light \(c\), and Boltzmann constant \(k\) are all set to 1.}
\label{fig1}
\end{figure}

Besides the MSC between Alice and Bob, we can also discuss the MSC between Alice and Anti-Bob, and the MSC between Bob and Anti-Bob. As Anti-Bob is inside of the event horizon, we use the term “inaccessible MSC”.

Tracing over the mode $B$, we can obtain
\begin{align}
\rho_{A\overline{B}}=\left(
                       \begin{array}{cccc}
                         \frac{1}{2} & 0 & 0 & -\frac{S}{2} \\
                         0 & 0 & 0 &0 \\
                         0 & 0 & \frac{C^2}{2} & 0 \\
                         -\frac{S}{2} & 0 & 0 & \frac{S^2}{2} \\
                       \end{array}
                     \right)
\end{align}

Following the calculation steps in above subsection, we can obtain the MSC between Alice and Anti-Bob, which is
\begin{align}
C(\rho_{A\overline{B}})=\frac{1}{\sqrt{1+e^{\frac{1}{T}}}},
\end{align}

Tracing over the mode A, we can obtain
\begin{align}
\rho_{B\overline{B}}=\frac{1}{2}\left(
                       \begin{array}{cccc}
                         C^2 & 0 & 0 & CS \\
                         0 & 0 & 0 & 0 \\
                         0 & 0 & 1 & 0 \\
                         CS & 0 & 0 & S^2 \\
                       \end{array}
                     \right)
\end{align}
The MSC between Bob and Anti-Bob are thus calculated as
\begin{align}
C(\rho_{B\overline{B}})=\frac{1}{2\sqrt{1+e^{-\frac{1}{T}}}\sqrt{1+e^{\frac{1}{T}}}\sqrt{1-\frac{1}{(1+e^{-\frac{1}{T}})^2}}}.
\end{align}

In Fig. \ref{fig1}, we plot the $C(\rho_{AB})$, $C(\rho_{A\overline{B}})$ and $C(\rho_{B\overline{B}})$ as functions of the Hawking temperature $T$ with $\omega=1$. We can see that (i)$C(\rho_{AB})$ is monotonically decreasing as the Hawking temperature increases and when $T\rightarrow\infty$, $C(\rho_{AB})$ is converge to $1/\sqrt{2}$; (ii)$C(\rho_{A\overline{B}})$ is monotonically increasing as the Hawking temperature increases and when $T\rightarrow\infty$, $C(\rho_{AB})$ is converge to $1/\sqrt{2}$; (iii) $C(\rho_{B\overline{B}})$ is monotonically increasing as the Hawking temperature increases. It has been shown that MSC has a close connection with discord-like correlations\cite{Hu2018}. The observed phenomena may be attributed to the redistribution of quantum correlations across different regions.

\section{Conclusions}\label{sec5}
In this study, we have theoretically examined and numerically simulated the effect of Hawking radiation on MSC in the Schwarzschild space-time. We have found that the accessible MSC decreases with an increase in Hawking temperature, while the inaccessible MSC shows a corresponding increase. This phenomenon suggests a redistribution of quantum correlations, previously accessible to inertial observers, across the bipartite modes due to the effects of Hawking radiation. Moreover,  It is interesting to note that, in limit case that the temperature tends to infinity, the accessible MSC equals to $1/\sqrt{2}$ of its initial
value, and the unaccessible MSC between mode $A$ and $\overline{B}$ also equals to the same value. Our results underscore the critical role of Hawking radiation in the evolution of quantum information properties near black holes and provide a new perspective on the behavior of quantum information in curved spacetime environments.

\begin{acknowledgments}
This work was supported by the Natural Science Research Start-up Foundation of Recruiting Talents of Nanjing University of Posts and Telecommunications (Grant No.NY222123) and Natural Science Foundation of Nanjing University of Posts and Telecommunications (Grant No.NY223069).
\end{acknowledgments}

%

\end{document}